%% file: output.tex
\documentclass[runningheads]{llncs}
\input{preamble}

\usepackage{orcidlink}
\usepackage{fullpage}
\usepackage[T1]{fontenc}

\usepackage{graphicx}

\begin{document}
\title{Solution Discovery for Vertex Cover, Independent Set, \\Dominating Set, and Feedback Vertex Set\thanks{This work was partially supported by JST SPRING Grant Number JPMJSP2114, the \textit{Baden-Württemberg-STIPENDIUM} from the Baden-Württemberg-Stiftung, and JSPS KAKENHI Grant Number JP25K21148.}}
\titlerunning{Solution Discovery for VC, IS, DS, and FVS}

\author{Rin Saito\inst{1}\orcidlink{0000-0002-3953-4339} \and
Anouk Sommer\inst{2}\orcidlink{0009-0006-1366-4377} \and
Tatsuhiro Suga\inst{1}\orcidlink{0009-0002-1376-4678} \and \\
Takahiro Suzuki\inst{1}\orcidlink{0009-0005-8433-3789} \and
Yuma Tamura \inst{1}\orcidlink{0009-0001-5479-7006}
}

\authorrunning{R. Saito et al.}

\institute{
Graduate School of Information Sciences, Tohoku University, Sendai, Japan\\
\email{\{rin.saito,suga.tatsuhiro.p5,takahiro.suzuki.q4\}@dc.tohoku.ac.jp, tamura@tohoku.ac.jp}
\and
Karlsruhe Institute of Technology, Germany\\
\email{anouk.sommer@student.kit.edu}
}
\maketitle               
\begin{abstract}
In the solution discovery problem for a search problem on graphs, we are given an initial placement of $k$ tokens on the vertices of a graph and asked whether this placement can be transformed into a feasible solution by applying a small number of modifications.
In this paper, we study the computational complexity of solution discovery for several fundamental vertex-subset problems on graphs, namely \prb{Vertex Cover Discovery}, \prb{Independent Set Discovery}, \prb{Dominating Set Discovery}, and \prb{Feedback Vertex Set Discovery}.
We first present XP algorithms for all four problems parameterized by clique-width. We then prove that \prb{Vertex Cover Discovery}, \prb{Independent Set Discovery}, and \prb{Feedback Vertex Set Discovery} are $\NP$-complete for chordal graphs and graphs of diameter~$2$, which have unbounded clique-width. In contrast to these hardness results, we show that all three problems can be solved in polynomial time on split graphs.
Furthermore, we design an FPT algorithm for \prb{Feedback Vertex Set Discovery} parameterized by the number of tokens.

\end{abstract}

\input{01_Introduction}
\input{02_Preliminaries}

\input{03_Parameter}
\input{04_GraphClass}
\input{05_conclusion}

\bibliographystyle{splncs04}
\bibliography{mybibliography}

\newpage
\appendix

\input{03_Parameter_appendix}
\input{04_GraphClass_appendix}

\end{document}

%% file: preamble.tex
\usepackage{graphicx}
\usepackage{esvect} 
\usepackage{tcolorbox}
\usepackage{complexity}
\usepackage{tikz}
\usepackage{amsmath}

\usepackage{amsthm}
\usepackage{amssymb}
\usepackage{amsfonts}
\usepackage{mathrsfs}
\usepackage{multibib}
\usepackage{enumerate}
\usepackage{algorithm}
\usepackage{algorithmic}
\usepackage{booktabs}
\usepackage{multirow}
\usepackage{tabularx}
\usepackage{arydshln}
\usepackage[mathlines]{lineno}
\usepackage{tabularx}
\usepackage{arydshln}
\usepackage{hyperref}
\usepackage{cleveref}
\DeclareMathOperator{\dist}{\textsf{dist}}
\DeclareMathOperator{\argmin}{arg\min}

\spnewtheorem{observation}{Observation}{\bfseries}{\itshape}

\newcommand{\prb}[1]{\textnormal{\textsc{#1}}}
\newcommand{\PiD}{${\rm \Pi}$-\prb{Discovery}}
\newcommand{\sourcePi}{${\rm \Pi}$}
\newcommand{\VCD}{\prb{VC-D}}
\newcommand{\FVSD}{\prb{FVS-D}}
\newcommand{\ISD}{\prb{IS-D}}
\newcommand{\DSD}{\prb{DS-D}}
\newcommand{\VC}{\prb{VC}}
\newcommand{\FVS}{\prb{FVS}}
\newcommand{\IS}{\prb{IS}}
\newcommand{\DS}{\prb{DS}}
\newcommand{\ECt}{\prb{Exact Cover by 3-Set}}

\newcommand{\true}{\texttt{true}}
\newcommand{\false}{\texttt{false}}
\newcommand{\conf}{S}

\newcommand{\yes}{\texttt{YES}}
\newcommand{\no}{\texttt{NO}}

\newcommand{\distribute}{\operatorname{\texttt{K}}}
\newcommand{\absorb}{\operatorname{\texttt{A}}}
\newcommand{\project}{\operatorname{\texttt{P}}}

\newcommand{\Cover}{C}
\newcommand{\Clique}{Q}
\newcommand{\cliq}{q}
\newcommand{\repres}{Y}

%% file: 01_Introduction.tex
\section{Introduction}\label{sec:Introduction}

Many conventional problems in algorithmic theory ask for a feasible solution from a given graph.
The feasible solution corresponds to a configuration of a real-world system, such as the placement of surveillance robots in a museum or factory.
However, in practical scenarios, a current configuration of a system often already exists, and the task is to obtain a feasible one starting from it. 
In particular, the current configuration may become infeasible due to system failures and must be restored to a feasible configuration as soon as possible.

To capture this type of situation, Fellows et al. proposed the framework of \emph{solution discovery}~\cite{SolDicov:FellowsGMMRRSS23}, and recently, several researchers have studied it in various problems~\cite{SolDicov:GroblerMMMRSS24,SolDicov:GroblerMMNRS24}.
In this framework, the problem {\PiD} is defined with respect to a combinatorial problem~${\rm \Pi}$ together with the upper bound of \emph{modification steps}, which specifies how many steps are allowed to modify a given current configuration into a feasible one. 
Formally, given an instance of~${\rm \Pi}$, an initial configuration $S$, and a budget $b$,  {\PiD} asks whether $S$ can be transformed into a feasible solution to ${\rm \Pi}$ within $b$ modification steps.  
Note that neither the initial nor the intermediate configurations are required to be feasible.
Most studies of solution discovery focus on vertex-subset problems on graphs, such as \prb{Vertex Cover} and \prb{Independent Set}, and an atomic modification step follows the \emph{token sliding model}~\cite{survey:Nishimura18}.
\footnote{We use the term ``token sliding'' following the terminology in the paper by Fellows et al.~\cite{SolDicov:FellowsGMMRRSS23} that introduced the solution discovery framework.}
In the token sliding model of {\PiD}, a vertex subset is regarded as a set of tokens placed on vertices, where each vertex is occupied by at most one token.
Then, at each step, one token may slide along an edge.
Recalling the example of placing surveillance robots, this model naturally represents the movement of robots between adjacent locations.

\subsection{Our Contribution}

This paper studies the computational complexity of \prb{Vertex Cover Discovery}, \prb{Independent Set Discovery}, \prb{Dominating Set Discovery}, and \prb{Feedback Vertex Set Discovery} ({\VCD}, {\ISD}, {\DSD}, and {\FVSD}, respectively) under the token sliding model. 
An overview of our results is provided in \Cref{tab:results}.\footnote{The \textsf{XNLP}-hardness of {\FVSD} parameterized by pathwidth follows by a simple reduction from {\VCD}; see \Cref{sec:parameterized}. 
}

The first three problems have been widely studied in previous work from the viewpoint of parameterized complexity~\cite{SolDicov:FellowsGMMRRSS23,SolDicov:GroblerMMNRS24}.
We first show that {\VCD}, {\ISD}, and {\DSD} all admit XP algorithms when parameterized by clique-width. 
Consequently, these problems can be solved in polynomial time on graphs of bounded clique-width, including graphs of bounded treewidth~\cite{cw:CourcelleO00}, cographs, and distance-hereditary graphs. 
In particular, this result extends the earlier result of Fellows et al. that provided XP algorithms for {\VCD}, {\ISD}, and {\DSD} parameterized by treewidth~\cite{SolDicov:FellowsGMMRRSS23}. 
We emphasize that this result is the best possible for clique-width: Grobler et al. proved the $\mathsf{XNLP}$-hardness for these problems parameterized by pathwidth~\cite{SolDicov:GroblerMMNRS24}, which implies that no FPT algorithm exists for clique-width unless $\FPT=\W[1]$, since graphs with bounded pathwidth have bounded treewidth, and then bounded clique-width~\cite{cw:CourcelleO00}.

We next turn to graph classes of unbounded clique-width, in particular, chordal graphs and their subclass, split graphs. 
We first observe that {\DSD} remains $\NP$-complete even on split graphs. 
This follows immediately from the fact that \prb{Dominating Set} is $\NP$-complete on split graphs~\cite{DSsplit:Bertossi84}. 
Remarkably, we prove that {\VCD} and {\ISD} are also $\NP$-complete on chordal graphs, even though their underlying problems \prb{Vertex Cover} and \prb{Independent Set} are solvable in linear time on chordal graphs~\cite{Chordal:Gavril72,chordal:Rose76}.
To complement this hardness result, we further show that {\VCD} and {\ISD} can be solved in polynomial time on split graphs.
From the positive results on split graphs (excluding {\DSD}) and cographs, which have small diameter, a natural research direction is to design polynomial-time algorithms for graphs of small diameter.
However, we observe that {\VCD}, {\ISD}, and {\DSD} are $\NP$-complete even for graphs of diameter~$2$.

In addition, this paper initiates the study of {\FVSD}.
{\FVS} is one of the most fundamental vertex-subset problems, appearing in the list of Karp's 21 NP-complete problems~\cite{Karp21/Karp10},
and has been extensively studied in parameterized complexity (see, e.g.,~\cite{PA:CyganFKLMPPS15}).
Following the approach for {\VCD}, we provide an XP algorithm for {\FVSD} parameterized by clique-width and a polynomial-time algorithm on split graphs, whereas {\FVSD} is shown to be $\NP$-complete on chordal graphs or graphs of diameter~$2$.
An interesting question is whether {\FVSD} admits an FPT algorithm when parameterized by the number of tokens $k$, since {\VCD}, {\ISD}, and {\DSD} are \W[1]-hard when parameterized by the feedback vertex set number~\cite{SolDicov:GroblerMMNRS24}, whereas {\VCD} is known to be fixed-parameter tractable when parameterized by~$k$~\cite{SolDicov:FellowsGMMRRSS23}.
Note that the FPT algorithm for {\VCD} in~\cite{SolDicov:FellowsGMMRRSS23} crucially relies on enumerating all minimal vertex covers of size at most $k$.
This strategy does not carry over to {\FVSD}, since the number of minimal feedback vertex sets of size at most $k$ can be as large as $O(n^k)$, and hence such enumeration cannot be done within FPT time. 
To overcome this obstacle, we exploit the concept of $k$-\emph{compact representations} of minimal feedback vertex sets~\cite{solsize:Guo06}, which yields a single-exponential FPT algorithm parameterized by $k$.

\Crefname{theorem}{Thm.}{Thms.}
\Crefname{corollary}{Cor.}{Cors.}
\begin{table}[t]
    \caption{The parameterized complexity and complexity on graph classes for {\VCD}, {\ISD}, {\DSD}, and {\FVSD}. The parameter $k$ denotes the number of tokens. The characters ``h'' and ``c'' represent ``hard'' and ``complete'', respectively.}
    \centering
    \begin{tabular}{ccccccc}
    \cline{4-7}
         \multicolumn{1}{c}{}&\multicolumn{1}{c}{}&\multicolumn{1}{c}{}&\multicolumn{4}{c}{\textbf{Problem}} \\ \cline{4-7}
         \multicolumn{1}{c}{}&\multicolumn{1}{c}{}&\multicolumn{1}{c}{}& {\VCD} & {\FVSD} & {\ISD} & {\DSD} \\ \hline\hline
         \multirow{3}{*}{\rotatebox{90}{\textbf{Para-}}}&\multirow{3}{*}{\rotatebox{90}{\textbf{meter}}}&$k$ & $\FPT$~\cite{SolDicov:FellowsGMMRRSS23} & $\FPT$~[\Cref{thm:FVSD_FPT_solsize}] & $\W[1]$-h~\cite{SolDicov:FellowsGMMRRSS23} & $\W[1]$-h~\cite{SolDicov:FellowsGMMRRSS23} \\ \cline{3-7}
         &&~clique-width~ & $\XP$~[\Cref{thm:algo_clique_width}] & $\XP$~[\Cref{thm:algo_clique_width}] & $\XP$~[\Cref{thm:algo_clique_width}] & $\XP$~[\Cref{thm:algo_clique_width}] \\ \cline{3-7}
         &&pathwidth & ~~$\mathsf{XNLP}$-h~\cite{SolDicov:GroblerMMNRS24}~~ & ~~$\mathsf{XNLP}$-h~~ & ~~$\mathsf{XNLP}$-h~\cite{SolDicov:GroblerMMNRS24}~~ & ~~$\mathsf{XNLP}$-h~\cite{SolDicov:GroblerMMNRS24}~~ \\ 
         \hline\hline
         \multirow{3}{*}{\rotatebox{90}{\textbf{Graph}}}&\multirow{3}{*}{\rotatebox{90}{\textbf{Class}}}&chordal& $\NP$-c~[\Cref{thm:VCD_NPcomp_chordal}] & $\NP$-c~[\Cref{cor:FVSD_NPcomp_chordal}] & $\NP$-c~[\Cref{cor:FVSD_NPcomp_chordal}] & $\NP$-c~\cite{DS_diam2:LokshtanovMPRS13} \\ \cline{3-7}
         &&diameter $2$ & $\NP$-c~[\Cref{cor:VCD_ISD_FVSD_NPcomp_diam}] & $\NP$-c~[\Cref{cor:VCD_ISD_FVSD_NPcomp_diam}] & $\NP$-c~[\Cref{cor:VCD_ISD_FVSD_NPcomp_diam}] & $\NP$-c~\cite{DS_diam2:LokshtanovMPRS13} \\ \cline{3-7}
         &&split & $\P$~[\Cref{thm:VCD_split}] & $\P$~[\Cref{cor:ISD_FVSD_polytime_split}] & $\P$~[\Cref{cor:ISD_FVSD_polytime_split}] & $\NP$-c~\cite{DS_diam2:LokshtanovMPRS13} \\ \hline
         
    \end{tabular}
    \label{tab:results}
\end{table}
\Crefname{theorem}{Theorem}{Theorems}
\Crefname{corollary}{Corollary}{Corollaries}

\subsubsection{Related Work}
Fellows et al. initiated the study of a solution discovery framework for several classical $\NP$-complete graph problems, including {\VC}, {\IS}, and {\DS}, under the token sliding model~\cite{SolDicov:FellowsGMMRRSS23}.
They analyzed the parameterized complexity with respect to various parameters, such as the configuration size $k$ and the budget $b$, and also considered it using notions of graph sparsity.
In subsequent work, Grobler et al.\ investigated the kernelization complexity of {\VCD}, {\ISD}, {\DSD}, and other problems under the token sliding model, and provided a refined classification with respect not only to $k$ and $b$, but also to structural parameters (e.g., pathwidth and feedback vertex set number).
In another line of research, Grobler et al.\ investigated solution discovery for polynomial-time solvable problems~\cite{SolDicov:GroblerMMMRSS24}.

Very recently, Bousquet et al.~\cite{SolDicov:BMMN25}
provided a meta-algorithm for solution discovery based on model checking of monadic second-order logic (MSO).
Note that their results do not imply an XP algorithm for clique-width, which is one of our contributions.

The token sliding model originates from research on \emph{combinatorial reconfiguration} problems (see the surveys by van den Heuvel~\cite{survey:Heuvel13} and Nishimura~\cite{survey:Nishimura18}),
which ask whether, given two feasible solutions of a combinatorial problem, there is a step-by-step transformation from one into another without losing the feasibility.
In contrast, solution discovery does not specify a target solution, and intermediate configurations may be infeasible.

\subsubsection*{Organization}
In \Cref{sec:Preliminaries}, we give preliminaries and define our problems.  
In \Cref{sec:parameterized}, we present XP algorithms for {\VCD}, {\ISD}, {\DSD}, and {\FVSD} parameterized by clique-width, and an FPT algorithm for {\FVSD} parameterized by the number of tokens.  
In \Cref{sec:NPcomp}, we analyze the complexity of {\VCD}, {\ISD}, and {\FVSD} on graph classes of unbounded clique-width, including chordal graphs, split graphs, and graphs of small diameter.  
The proofs of claims marked $(\ast)$ can be found in the Appendix.

%% file: 02_Preliminaries.tex
\section{Preliminaries}\label{sec:Preliminaries}
For a positive integer~$p$, we write $[p] = \{1,2,\ldots,p\}$.
Let $G = (V, E)$ be an undirected graph, and let $V(G)$ and $E(G)$ denote its vertex set and edge set, respectively. 
For a vertex $v \in V(G)$, we define the \emph{open neighborhood} of $v$ as $N(v) = \{u \in V(G) \colon uv \in E(G)\}$, and the \emph{closed neighborhood} as $N[v] = N(v) \cup \{v\}$.
For vertices $u,v \in V(G)$, we write $\dist(u,v)$ for the number of edges in a shortest path between $u$ and $v$.
The \emph{diameter} of a connected graph $G$, denoted by $ \mathsf{diam}(G)$, is $\max_{u,v\in V}\dist(u,v)$.

A \emph{vertex cover} of $G$ is a set $C \subseteq V(G)$ such that every edge has an endpoint in $C$.   
An \emph{independent set} of $G$ is a set $I \subseteq V(G)$ such that no two vertices in $I$ are adjacent.  
A \emph{dominating set} of $G$ is a set $D \subseteq V(G)$ such that every vertex of $G$ belongs to $D$ or has a neighbor in $D$.   
A \emph{feedback vertex set} of $G$ is a set $F \subseteq V(G)$ such that every cycle of $G$ contains at least one vertex from $F$.  

Given a graph $G$ and an integer $k$, the problems \prb{Vertex Cover} ({\VC}), \prb{Independent Set} ({\IS}), \prb{Dominating Set} ({\DS}), and \prb{Feedback Vertex Set} ({\FVS}) ask whether $G$ contains such a set of size at most $k$ (for {\VC}, {\DS}, {\FVS}) or at least $k$ (for {\IS}).

A \emph{split graph} is a graph whose vertex set can be partitioned into a clique and an independent set.  
A \emph{chordal graph} is a graph in which every cycle of length at least $4$ has a chord, i.e., an edge between two non-consecutive vertices of the cycle.

\subsubsection{Our problems}
Let $G$ be a graph.  
In this work, a \emph{configuration} of $G$ is simply a vertex subset of $G$.
Two configurations $\conf$ and $\conf'$ of $G$ are \emph{adjacent} if there exist $x \in \conf$ and $y \in V(G)\setminus \conf$ with $xy \in E(G)$ such that $\conf' = \conf \cup \{y\} \setminus \{x\} $. (In particular, $\lvert \conf \rvert = \lvert \conf' \rvert$ holds.)
A configuration $\conf_t$ is \emph{reachable in $\ell$ steps} from $\conf_s$ if there exists a sequence of configurations $\langle \conf_s = \conf_0,\conf_1,\conf_2,\ldots \conf_\ell=\conf_t\rangle$ such that $\conf_{i-1}$ and $\conf_i$ are adjacent for all $i \in [\ell]$.
We now define four discovery problems: \prb{Vertex Cover Discovery} ({\VCD}), \prb{Independent Set Discovery} ({\ISD}), \prb{Dominating Set Discovery} ({\DSD}), and \prb{Feedback Vertex Set Discovery} ({\FVSD}). 
Given a graph $G$, an initial configuration $\conf$ of size $k$, and a non-negative integer $b$ (called the \emph{budget}), the problem {\VCD} ({\ISD}, {\DSD}, and {\FVSD}, respectively) asks whether there exists a configuration $\conf_t$ that forms a vertex cover (independent set, dominating set, and feedback vertex set, respectively) of $G$ and is reachable from $\conf$ in at most $b$ steps.

Throughout this paper, we assume that the input graph is connected; otherwise, the problems can be solved independently on each connected component. 
We also assume $b \leq k \cdot \mathsf{diam}(G)$ because each token can traverse a path of length at most $\mathsf{diam}(G)$.  
Moreover, when $b = k \cdot \mathsf{diam}(G)$, the problem {\PiD} is equivalent to the corresponding underlying problem {\sourcePi}: given an instance $(G,k)$ of {\sourcePi}, we construct an instance $(G,S,b)$ of {\PiD}, where $S$ is an arbitrary vertex-subset of size $k$ and $b = k \cdot \mathsf{diam}(G)$.  
Since every configuration of size $k$ is reachable from $S$ within $b$ steps, $(G,k)$ is a yes-instance of {\sourcePi} if and only if $(G,S,b)$ is a yes-instance of {\PiD}.  

Finally, note that all four problems are in $\NP$. 
Therefore, to prove their $\NP$-completeness, it suffices to show their $\NP$-hardness.

%% file: 03_Parameter.tex
\section{Parameterized Algorithms} \label{sec:parameterized}
In this section, we investigate the parameterized complexity of {\VCD}, {\ISD}, {\DSD}, and {\FVSD}.

It is known that {\VCD}, {\ISD}, and {\DSD} are $\mathsf{XNLP}$-hard when parameterized by pathwidth~\cite{SolDicov:GroblerMMNRS24}, which implies that no FPT algorithm exists for pathwidth unless $\FPT=\W[1]$.
We further observe that {\FVSD} is also $\mathsf{XNLP}$-hard when parameterized by pathwidth: given an instance of {\VCD}, we can construct an equivalent instance of {\FVSD} by replacing each edge $uv$ of the input graph with a triangle on vertices $u, v, e_{uv}$. 
This transformation increases the pathwidth by exactly one (see also the definition of pathwidth in~\cite{PA:CyganFKLMPPS15}), 
and one can observe that this yields a parameterized logspace reduction (see the definition in~\cite{pl:BODLAENDER2024}) from {\VCD} to {\FVSD}.

In contrast to these hardness results, we show that all four problems admit XP algorithms when parameterized by clique-width. 
Recall that the clique-width of a graph is always bounded from above by its treewidth and pathwidth. 
Therefore, our result strictly strengthens the XP algorithm of Fellows et al.~\cite{SolDicov:FellowsGMMRRSS23} for {\VCD}, {\ISD}, and {\DSD} parameterized by treewidth.

Finally, we show that {\FVSD} admits a single-exponential FPT algorithm when parameterized by the number of tokens $k$.
Note that the one for {\VCD} is already known, and {\ISD} and {\DSD} are $\W[1]$-hard when parameterized by $k+b$~\cite{SolDicov:FellowsGMMRRSS23}.

\subsection{XP algorithm parameterized by clique-width}\label{sec: FPT/cw}
In this subsection, we first present an XP algorithm for {\VCD} parameterized by clique-width, based on dynamic programming on a tree structure known as a \emph{$w$-expression}.
Then, by applying the same framework, we also show XP algorithms for {\FVSD}, {\ISD}, and {\DSD} parameterized by clique-width.

We now give the definition of the clique-width of graphs.
For a positive integer $w$, a \emph{$w$-graph} is a graph in which each vertex is assigned a label from $\{1, 2, \dots, w\}$.
We define the following four operations on $w$-graphs.
\begin{description}
    \item[Introduce $i(v)$:] Create a graph consisting of a single vertex $v$ labeled with $i \in [w]$.
    \item[Union $\oplus$:] Take the disjoint union of two $w$-graphs.
    \item[Relabel $\rho_{i \to j}$:] For distinct $i, j \in [w]$, replace every label $i$ in the graph with $j$.
    \item[Join $\eta_{i, j}$:] For distinct $i, j \in [w]$, add all possible edges between vertices labeled $i$ and vertices labeled $j$.
\end{description}

The \emph{clique-width} of a graph $G$ is the smallest integer $w$ such that $G$ can be constructed from $w$-graphs by a sequence of these operations.
Such a construction can be represented by a rooted tree, where each node corresponds to one of the above operations. 
This tree is called a \emph{$w$-expression tree} of $G$, and its nodes are referred to as \emph{introduce}, \emph{union}, \emph{relabel}, or \emph{join} nodes, respectively. 
For a node $t$ of the $w$-expression tree $T$, let $G_t$ denote the labeled graph associated with the subtree rooted at $t$. 
For node $t$ of $T$ and $i \in [w]$, we call a vertex of $G_t$ with label $i$ an \emph{$i$-vertex}, and $U^t_i$ denotes the set of $i$-vertices in $G_t$.
It is known that, given a graph $G$ of clique-width $\mathrm{cw}$, one can compute a $(2^{\mathrm{cw}+1}-1)$-expression of $G$ in $O(|V(G)|^3)$ time~\cite{cw:HlinenyO08,cw:Oum08,cw:OumS06}. 
A $w$-expression tree is called \emph{irredundant} if, for each edge $uv$ of $G$, there exists exactly one join node $\eta_{i,j}$ that introduces the edge $uv$. 
Moreover, any $w$-expression tree can be transformed in linear time into an irredundant $w$-expression tree with $O(n)$ nodes~\cite{cw:CourcelleO00}. 
Hence, in the following, we may always assume without loss of generality that the $w$-expression tree is irredundant.

\subsubsection{Dynamic Programming}
Let $(G,S,b)$ be an instance of {\VCD}, where $G$ has $n$ vertices and $T$ is a $w$-expression tree of $G$.  
We perform a bottom-up dynamic programming over $T$, where the update rules depend on the type of each node in $T$. 

To simplify the description of our dynamic programming, we introduce a hypothetical vertex $v^*$ that can hold an arbitrary number of tokens and is adjacent to every vertex of $G_t$ for each node $t$ of $T$. 
Intuitively, the vertex $v^*$ virtually represents all vertices of $G$ ``outside'' $G_t$.
We then place $k - |V(G_t) \cap S|$ tokens on $v^*$ in the initial configuration. 
Let $G^*_t$ denote the graph obtained from $G_t$ by adding $v^*$ and connecting it to all vertices of $G_t$.

For each node $t$ of $T$, we maintain a Boolean DP table indexed by tuples $s = (\ell, \distribute, \absorb, \project)$, where $\distribute, \absorb, \project$ are functions such that $\distribute \colon [w] \to [k]$ and $\absorb, \project \colon [w] \to [b]$.
Here, $\ell$ denotes the number of token moves used in $G_t$ ($0 \le \ell \le b$). 
For each label $i \in [w]$, $\distribute(i)$ represents the number of tokens placed on vertices with label $i$ as the final configuration in $G^*_t$. 
The values $\absorb(i)$ and $\project(i)$ indicate, respectively, the numbers of incoming and outgoing additional moves (\emph{absorption} and \emph{projection}) between $i$-vertices and $v^*$ in $G^*_t$. 
Note that, for each node $t$, the number of possible tuples is at most $b \cdot k^w \cdot b^{2w} \le n^{5w+2}$ since $k \le n$ and $b \le k \cdot \mathsf{diam}(G) \le n^2$.

For each tuple $s$, the table entry $c_t[s]$ is set to $\true$ if and only if there exists a vertex cover $C$ of $G_t$ and a sequence $\mathcal{S}$ of configurations in $G^*_t$ such that
(i) $|C \cap U^t_z| = \distribute(z)$ for every $z \in [w]$;  
(ii) $\mathcal{S}$ performs exactly $\ell$ token moves within $G_t$, with $\ell \leq b$;  
(iii) exactly $\absorb(z)$ moves are performed from $v^*$ to the vertices in $U^t_z$; and  
(iv) exactly $\project(z)$ moves are performed from the vertices in $U^t_z$ to $v^*$. 

A tuple $s = (\ell, \distribute, \absorb, \project)$ is said to be \emph{valid} if it satisfies the following conditions:  
(a) $\sum_{z \in [w]} \distribute(z) \le k$;  
(b) if $|U_z^t| = 0$ for $z \in [w]$, then $\distribute(z) = \absorb(z)=\project(z) = 0$;
(c) $\ell \le b$; and 
(d) $|S \cap U^t_z| + \absorb(z) - \project(z) = \distribute(z) \leq |U^t_z|$ for all $z \in [w]$. 
Intuitively, condition~(a) ensures that the number of tokens in $G_t$ does not exceed $k \ (=|S|)$.
Condition~(b) states that if $U_z^t$ is empty, then there is no token on $z$-vertices as a final configuration, and no token moves occur between $z$-vertices and $v^*$. 
Condition~(c) guarantees that the total number of token moves occurring within $G_t$ does not exceed the budget $b$. Condition~(d) enforces the consistency of the number of tokens between the initial and final configurations. 
Note that, since $|S \cap U^t_z|$ and $|U^t_z|$ are already known for all $z \in [w]$, one can check whether a tuple $s$ is valid in $O(w)$ time.
At the root node $r$ of $T$, our algorithm outputs $\yes$ if and only if there exists a valid tuple $s$ with $c_r[s] = \true$ such that $\absorb(z) = \project(z) = 0$ for all $z \in [w]$. 
As we will see later, the DP entry corresponding to any invalid tuple is always assigned the value $\false$, since such moves cannot be realized.

\smallskip
\noindent\textbf{Introduce node $i(v)$:}
Let $t$ be an introduce node of $T$, and let $s = (\ell, \distribute, \absorb, \project)$ be a tuple for $t$.  
In this case, the subgraph $G_t$ consists only of a single vertex $v$. 
Since both the empty set $\emptyset$ and singleton $\{v\}$ form vertex covers of $G_t$, we set $c_t[s] = \true$ if and only if $s$ is a valid tuple and $\ell = 0$.

\smallskip
\noindent\textbf{Union node $\oplus$:}
Let $t$ be a union node of $T$ with children $t_1$ and $t_2$, and let $s = (\ell, \distribute, \absorb, \project)$ be a tuple for $t$. 
As $G_t$ is the disjoint union of $G_{t_1}$ and $G_{t_2}$, no tokens move between them.
Recall that the union of a vertex cover of $G_{t_1}$ and a vertex cover of $G_{t_2}$ forms a vertex cover of $G_t$. 
Thus, each value of $\ell, \distribute(i), \absorb(i), \project(i)$ ($i \in [w]$) must be the sum of the corresponding values for $t_1$ and $t_2$, respectively.
Consequently, we set $c_t[s] = \true$ if and only if $s$ is valid and there exist tuples $s_1 = (\ell_1, \distribute_1, \absorb_1, \project_1)$ for $t_1$ and $s_2 = (\ell_2, \distribute_2, \absorb_2, \project_2)$ for $t_2$ such that
\begin{enumerate}
    \item $c_{t_1}[s_1] = \true$ and $c_{t_2}[s_2] = \true$,
    \item $\ell = \ell_1 + \ell_2$, and
    \item $\distribute(z) = \distribute_1(z) + \distribute_2(z)$, $\absorb(z) = \absorb_1(z) + \absorb_2(z)$, and $\project(z) = \project_1(z) + \project_2(z)$ for all $z \in [w]$.
\end{enumerate}

\smallskip
\noindent\textbf{Relabel node $\rho_{i \to j}$:}
Let $t$ be a relabel node of $T$ with child $t'$, and let $s = (\ell, \distribute, \absorb, \project)$ be a tuple for $t$. 
Since all $i$-vertices of $G_{t'}$ are relabeled as $j$-vertices in $G_t$ (i.e., $U^t_i = \emptyset$), we must have $\distribute(i) = \absorb(i) = \project(i) = 0$, and the values of $i$-vertices for $t'$ are merged into those of $j$-vertices.  
Consequently, we set $c_t[s] = \true$ if and only if $s$ is valid, $\distribute(i) = \absorb(i) = \project(i) = 0$, and there exists a valid tuple $s' = (\ell', \distribute', \absorb', \project')$ for $t'$ such that
\begin{enumerate}
    \item $c_{t'}[s'] = \true$,
    \item $\ell = \ell'$,
    \item $\distribute(j) = \distribute'(i) + \distribute'(j)$, $\absorb(j) = \absorb'(i) + \absorb'(j)$, $\project(j) = \project'(i) + \project'(j)$, and
    \item $\distribute(z) = \distribute'(z)$, $\absorb(z) = \absorb'(z)$, $\project(z) = \project'(z)$ for all $z \in [w] \setminus \{i,j\}$.
\end{enumerate}

\smallskip
\noindent\textbf{Join node $\eta_{i,j}$:}
Let $t$ be a join node of $T$ with child $t'$, and let $s = (\ell, \distribute, \absorb, \project)$ be a tuple for $t$.
Since $T$ is irredundant, all edges between $i$-vertices and $j$-vertices are added in $G_t$. 
Thus, in any vertex cover $C$ of $G_t$, at least one of the equalities $\distribute(i) = |U^t_i \cap C|$ or $\distribute(j) = |U^t_j \cap C|$ must hold; otherwise, some edge between $U_i^t$ and $U_j^t$ is uncovered.
Therefore, if both $\distribute(i) < |U^t_i|$ and $\distribute(j) < |U^t_j|$, then we set $c_t[s] = \false$.  

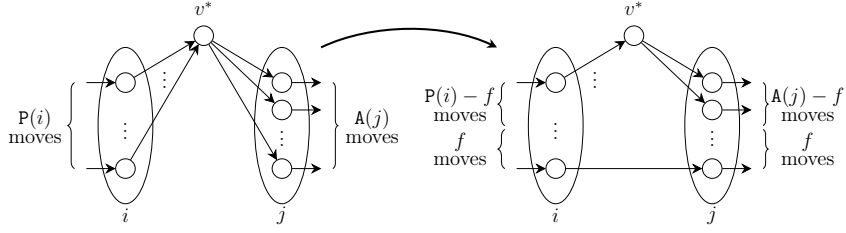
\begin{figure}[t]
    \centering
    \scalebox{0.65}{\input{figures/cw_join}}
    \caption{A rough illustration for a join node. If $f$ tokens directly move from the $i$-vertices to the $j$-vertices, we decrease both $\project(i)$ and $\absorb(j)$ by $f$, and increase the total number of moves $\ell$ by $f$.}
    \label{fig:cwjoin}
\end{figure}

Otherwise (i.e., $\distribute(i) = |U^t_i|$ or $\distribute(j) = |U^t_j|$), we consider the situation that some tokens move between $i$-vertices and $j$-vertices. 
Let $f \ge 0$ denote the number of additional moves from $i$-vertices to $j$-vertices, and let $g \ge 0$ denote the number of additional moves in the opposite direction. 
In other words, we allow tokens to move directly between $i$-vertices and $j$-vertices, which were previously treated as moves from $i$-vertices to $j$-vertices (or opposite direction), passing through $v^*$.
Therefore, we then set $c_t[s] = \true$ if and only if $s$ is valid and there exist integers $f, g \ge 0$ together with a valid tuple $s' = (\ell', \distribute', \absorb', \project')$ for $t'$ such that
\begin{enumerate}
    \item $c_{t'}[s'] = \true$,
    \item $\ell = \ell' + f + g$,
    \item $\distribute(z) = \distribute'(z)$ for all $z \in [w]$,
    \item $\absorb(z) = \absorb'(z)$ and $\project(z) = \project'(z)$ for all $z \in [w] \setminus \{i,j\}$, and
    \item $\absorb(i) = \absorb'(i) - g \geq 0$, $\absorb(j) = \absorb'(j) - f \geq 0$, $\project(i) = \project'(i) - f \geq 0$, $\project(j) = \project'(j) - g \geq 0$.
\end{enumerate}
Regarding Conditions~2 and~5, note that there are $f$ tokens that were moving from the $i$-vertices to the $j$-vertices via $v^{\ast}$, and $g$ tokens that were moving from the $j$-vertices to the $i$-vertices via $v^{\ast}$. 
Since they originally passed through $v^{\ast}$ without incurring any moves, condition 2 makes this correction.
In condition~5, we distribute the tokens that were passing through $v^{\ast}$ at node $t'$.
A brief illustration is given in \Cref{fig:cwjoin}.

\smallskip
\noindent\textbf{Running time.}
Observe that the validity of a given tuple $s$ can be verified in $O(w)$ time, and for each type of node $t$, the value $c_t[s]$ can also be computed within $n^{O(w)}$ time. 
Since the DP table of a node $t$ contains at most $n^{5w+2}$ entries and the expression tree $T$ has $O(n)$ nodes, {\VCD} can be solved in $n^{O(w)}$ time.

\smallskip
\noindent\textbf{Remark.}
The central idea of our algorithm is to perform dynamic programming over a $w$-expression by guessing the two pieces of information: the feasibility of solutions and the number of token movements. 
More precisely, we guess the number of tokens on each $i$-vertex ($i \in [w]$) in the final configuration for {\VCD}, and furthermore, we maintain auxiliary information that records the number of token movements between $i$-vertices and $v^*$ in $G^*_t$, specifically $\ell$, $\absorb$, and $\project$.

In other solution discovery problems, the first part of our approach can be inherited from the well-established dynamic programming approach for the underlying search problem parameterized by clique-width, which has already been known for {\FVS}~\cite{FVS:BergougnouxK19,FVS:Bui-XuanSTV13}, {\IS}, and {\DS}.
For instance, to solve {\FVSD}, one additionally keeps track of information for the connectivity constraints. 
Consequently, the same dynamic programming technique as {\VCD} can also be applied to {\FVSD}, {\ISD}, and {\DSD}. 
This yields the following result.

\begin{theorem}
    \label{thm:algo_clique_width}
    Given a $w$-expression tree of the input graph, the problems {\VCD}, {\FVSD}, {\ISD}, and {\DSD} can be solved in $n^{O(w)}$ time.
\end{theorem}

\subsection{FPT algorithm parameterized by the number of tokens}\label{subsec:FVS_FPT_k}

This section presents an FPT algorithm for $\FVSD$ parameterized by the number of tokens $k$.

\begin{theorem}\label{thm:FVSD_FPT_solsize}
    {\FVSD} parameterized by the number of tokens $k$ can be solved in $O(23.1^k n^3)$ time, where $n$ denotes the number of vertices in the input graph.
\end{theorem}

Our algorithm uses an approach for {\VCD} provided in~\cite{SolDicov:FellowsGMMRRSS23}.
We first generate all candidate solutions to {\FVS}.
Then, for each candidate solution, construct an edge-weighted complete bipartite graph $H$ and find a minimum weight matching of $H$ saturating one side of $H$.
The key idea is that if one could enumerate all inclusion-wise minimal feedback vertex sets of size at most $k$, then the above approach yields an FPT algorithm for $k$.
Unfortunately, in general, enumerating minimal feedback vertex sets cannot be done in FPT time with respect to $k$.
For example, consider a graph consisting of $k$ vertex-disjoint cycles of length $n/k$.
The graph contains $\mathrm{\Omega}((n/k)^k)$ distinct minimal feedback vertex sets.

We overcome this difficulty by utilizing the concept of \textit{compact representations} of minimal feedback vertex sets~\cite{solsize:Guo06}.
The family $\mathcal{\repres}=\{\repres_1,\repres_2,\dots\}$ of pairwise disjoint vertex subsets of a graph $G$ is a \emph{compact representation} of minimal feedback vertex sets if every set $F$ with $|F\cap \repres|=1$ for all $\repres\in \mathcal{\repres}$ forms a minimal feedback vertex set of $G$.
In particular, if $|\mathcal{\repres}|\le k$, then $\mathcal{\repres}$ is called a \emph{$k$-compact representation.}
We say that $\mathcal{\repres}$ \emph{represents} a minimal feedback vertex set $F$ of $G$ if $|F\cap \repres|=1$ for every $\repres\in \mathcal{\repres}$.
A list $\mathfrak{L}$ of $k$-compact representations is said to be \textit{complete} if, for every minimal feedback vertex set $F$ of size at most $k$, there is $\mathcal{\repres} \in \mathfrak{L}$ that represents $F$.
In other words, every feedback vertex set of size at most $k$ is represented by at least one compact representation in $\mathfrak{L}$.

We use the following result as a subroutine for our algorithm.
\begin{theorem}[Theorem 3 of \cite{solsize:Misra12}] \label{the:compact_rep}
    Given a graph $G$ with $m$ edges and a positive integer $k$, there exists an algorithm that computes a complete list of $k$-compact representations of $G$ in $O(23.1^k m)$ time.
    Moreover, the number of k-compact representations produced by the algorithm is at most $O(23.1^k)$.
\end{theorem}

Let $(G, S, b)$ be an instance of {\FVSD}.
In our algorithm, we first construct a complete list $\mathfrak{L}$ of $k$-compact representations of minimal feedback vertex sets in $G$ by \Cref{the:compact_rep}.
Second, for each $k$-compact representation $\mathcal{\repres} \in \mathfrak{L}$, we examine whether there exists at least one feedback vertex set represented by $\mathcal{\repres}$ that is reachable from $S$ within $b$ steps.

Now, we explain the second step of our algorithm.
Fix a $k$-compact representation $\mathcal{\repres} \in \mathfrak{L}$. 
We construct an edge-weighted complete bipartite graph 
$H_\mathcal{\repres}$ with bipartitions $S$ and $\mathcal{\repres}$.
To distinguish the vertices of $H_\mathcal{\repres}$ and the vertices of $G$, we refer to the vertices of $H_\mathcal{\repres}$ as \emph{nodes}.
For $u\in S$ and $\repres\in \mathcal{\repres}$, we assign the weight of the edge $(u,\repres)$ to $\min_{y\in \repres}\dist(u,y)$, 
which corresponds to the minimum number of steps required to move the token initially placed on $u$ to some vertex in $\repres$. 
We then consider a minimum-weight matching $M^*$ of $H_\mathcal{\repres}$ such that each node in $\mathcal{\repres}$ of $H_\mathcal{\repres}$ is incident to some edge in $M^*$.

The following \Cref{lem:solsizeFPT} establishes the correspondence between the total weight of $M$ and the minimum number of steps to reach some minimal feedback vertex sets represented by $\mathcal{Y}$ from $S$.
The proof can be found in \Cref{appx_subsec:solsizeFPT}.
\begin{lemma}[$\ast$]\label{lem:solsizeFPT}
The two statements are equivalent:
\begin{enumerate}
    \item[$(\mathrm{1})$] There exists a matching $M=\{(u_1,\repres_1),(u_2,\repres_2),\ldots, (u_{|\mathcal{\repres}|}, \repres_{|\mathcal{\repres}|})\}$ of $H_\mathcal{\repres}$ with total weight at most $b$, where $u_i \in S$ and $Y_i \in \mathcal{\repres}$ for $i \in [|\mathcal{\repres}|]$; and 
    \item[$(\mathrm{2})$] There exists a feedback vertex set $F$ that is reachable in $b$ steps from $S$ and a minimal feedback vertex set $F' \subseteq F$ represented by $\mathcal{\repres}$.
\end{enumerate}
\end{lemma}

Our algorithm outputs $\yes$ if there exists a $k$-compact representation $\mathcal{\repres} \in \mathfrak{L}$ such that the corresponding bipartite graph $H_{\mathcal{\repres}}$ admits a minimum-weight matching $M^*$ of total weight at most $b$, and outputs $\no$ otherwise.

Finally, we analyze the running time of our algorithm.
A complete list $\mathfrak{L}$ of $k$-compact representations such that $|\mathfrak{L}| = O(23.1^k)$ can be obtained in $O(23.1^k m)$ time~\cite{solsize:Misra12}.  
To construct the bipartite graph $H_\mathcal{\repres}$ for $\mathcal{\repres}\in \mathfrak{L}$, we precompute in $O(k(n+m))$ time the distances from every pair of vertices $u\in S$ and $v\in V$ by applying a breadth-first search algorithm $k$ times.
For each $\mathcal{\repres}\in \mathfrak{L}$, the graph $H_\mathcal{\repres}$ is constructed in $O(kn)$ time; for each $u \in S$, the weights of the edges incident to $u$ can be decided in $O(n)$ time because $\sum_{\mathcal{\repres}\in \mathfrak{L}} |\mathcal{\repres}| = O(n)$.
Since $H_\mathcal{\repres}$ has at most $2k$ vertices, the desired minimum-weight matching $M^*$ of $H_\mathcal{\repres}$ can be obtained in $O(|V(H_\mathcal{\repres})|^3) = O(k^3)$ time using the Hungarian algorithm~\cite{solsize:MWPM}.
Therefore, the overall running time of our algorithm is $O(23.1^km + k(n+m) + 23.1^k(kn + k^3))=O(23.1^k(n^2+k^3))=O(23.1^kn^3)$.

%% file: figures/cw_join.tex
\begin{tikzpicture}[scale =0.8, >=stealth]
\centering
\usetikzlibrary{positioning}
\usetikzlibrary{decorations.pathreplacing}
\usetikzlibrary{arrows.meta}
\tikzset{every node/.style={font=\large}}
\tikzset{
  >={Stealth[length=2mm,width=2mm]}
}
  \draw (-2,0) ellipse [x radius=0.7, y radius=2];
  \node[align = center] at (-2,-2.3) {$i$};
  \draw ( 2,0) ellipse [x radius=0.7, y radius=2];
  \node[align = center] at (2,-2.3) {$j$};

    \node[draw=black, circle, minimum size=4mm, inner sep=0pt] (P) at (0,2.3) {};
   \node[align=center,above=1mm of P]
    {$v^{\ast}$};
  \node[draw=black, circle, minimum size=4mm, inner sep=0pt] (i1) at (-2,1.1) {};
  \node[draw=black, circle, minimum size=4mm, inner sep=0pt] (i2) at (-2,-1.1) {};
  \draw[->] (-3,1.1) -- (i1);
  \draw[->] (-3,-1.1) -- (i2);
  \draw[->] (i1) -- (P);
  \draw[->] (i2) -- (P);
  \node at (-2,0) {$\vdots$};
  \node at (-1,1.3) {$\vdots$};
  \draw[
    decorate,
    decoration={brace,amplitude=4pt}  
  ] (-3.3,-1.1) -- (-3.3,1.1)
    node[midway,xshift=-23pt]{\shortstack{$\project(i)$\\moves}};

  \node[draw=black, circle, minimum size=4mm, inner sep=0pt] (j1) at (2,1.1) {};
  \node[draw=black, circle, minimum size=4mm, inner sep=0pt] (j3) at (2,0.4) {};
  \node[draw=black, circle, minimum size=4mm, inner sep=0pt] (j2) at (2,-1.1) {};
  \draw[->] (P) -- (j1);
  \draw[->] (P) -- (j2);
  \draw[->] (P) -- (j3);
  \draw[->] (j1) -- (3,1.1);
  \draw[->] (j2) -- (3,-1.1);
  \draw[->] (j3) -- (3,0.4);
  \node at (2,-0.2) {$\vdots$};
  \draw[
    decorate,
    decoration={brace,amplitude=3pt}  
  ] (3.3,1.1) -- (3.3,-1.1)
    node[midway,xshift=23pt]{\shortstack{$\absorb(j)$\\moves}};

\draw[->, very thick]
    (3,2) arc[
      start angle=130,
      end angle=50,
      x radius=3.5cm,
      y radius=2cm
    ];

\begin{scope}[xshift=11cm]
  \draw (-2,0) ellipse [x radius=0.7, y radius=2];
  \node[align = center] at (-2,-2.3) {$i$};
  \draw ( 2,0) ellipse [x radius=0.7, y radius=2];
  \node[align = center] at (2,-2.3) {$j$};

    \node[draw=black, circle, minimum size=4mm, inner sep=0pt] (P) at (0,2.3) {};
   \node[align=center,above=1mm of P]
    {$v^{\ast}$};
  \node[draw=black, circle, minimum size=4mm, inner sep=0pt] (i1) at (-2,1.1) {};
  \node[draw=black, circle, minimum size=4mm, inner sep=0pt] (i2) at (-2,-1.1) {};
  \draw[->] (-3,1.1) -- (i1);
  \draw[->] (-3,-1.1) -- (i2);
  \draw[->] (i1) -- (P);
  \node at (-2,0) {$\vdots$};
  \node at (-1,1.3) {$\vdots$};
  \draw[
    decorate,
    decoration={brace,amplitude=4pt}  
  ] (-3.3,0) -- (-3.3,1.1)
    node[midway,xshift=-26pt]{\shortstack{$\project(i)-f$\\moves}};
    \draw[
    decorate,
    decoration={brace,amplitude=4pt}  
  ] (-3.3,-1.1) -- (-3.3,-0.1)
    node[midway,xshift=-26pt]{\shortstack{$f$\\moves}};

  \node[draw=black, circle, minimum size=4mm, inner sep=0pt] (j1) at (2,1.1) {};
  \node[draw=black, circle, minimum size=4mm, inner sep=0pt] (j3) at (2,0.4) {};
  \node[draw=black, circle, minimum size=4mm, inner sep=0pt] (j2) at (2,-1.1) {};
  \draw[->] (P) -- (j1);
  \draw[->] (P) -- (j3);
  \draw[->] (j1) -- (3,1.1);
  \draw[->] (j2) -- (3,-1.1);
  \draw[->] (j3) -- (3,0.4);
  \draw[->] (i2) -- (j2);
  \node at (2,-0.2) {$\vdots$};
  \draw[
    decorate,
    decoration={brace,amplitude=3pt}  
  ] (3.3,1.1) -- (3.3,0)
    node[midway,xshift=26pt]{\shortstack{$\absorb(j)-f$\\moves}};
    \draw[
    decorate,
    decoration={brace,amplitude=3pt}  
  ] (3.3,-0.1) -- (3.3,-1.1)
    node[midway,xshift=26pt]{\shortstack{$f$\\moves}};
\end{scope}
\end{tikzpicture}

%% file: 04_GraphClass.tex
\section{Graphs of Unbounded Clique-width}\label{sec:NPcomp}
In this section, we analyze the computational complexity of {\VCD}, {\ISD}, and {\FVSD} on chordal graphs, split graphs, and graphs of bounded diameter, which have unbounded clique-width.
In \Cref{subsec:NP_comp}, we establish the $\NP$-completeness of these problems on chordal graphs and graphs of diameter $2$.
In \Cref{subsec:polytime_algo}, we show the polynomial-time solvability of these problems on split graphs.
We remark that {\DSD} is $\NP$-complete (more precisely, $\W[2]$-hard for the number of tokens $k$) for split graphs of diameter $2$, which follows directly from the known result that {\DS} is $\NP$-complete on those graphs~\cite{DS_diam2:LokshtanovMPRS13}.

\subsection{$\NP$-completeness}\label{subsec:NP_comp}
We show the $\NP$-completeness of {\VCD}, {\ISD}, and {\FVSD} for chordal graphs and graphs of diameter $2$.
We first show the following theorem. 
\begin{theorem}\label{thm:VCD_NPcomp_chordal}
    {\VCD} is $\NP$-complete for chordal graphs.
\end{theorem}

For the proof of \Cref{thm:VCD_NPcomp_chordal}, we reduce the {\ECt} problem to {\VCD}.  
In {\ECt}, we are given a ground set $U$ of $3n$ elements together with a family $\mathcal{X}$ of three-element subsets of $U$.  
The task is to decide whether there exists a subfamily of $\mathcal{X}$ consisting of exactly $n$ sets whose union forms $U$. 
It is known that {\ECt} is $\NP$-complete~\cite{GareyJ79}.

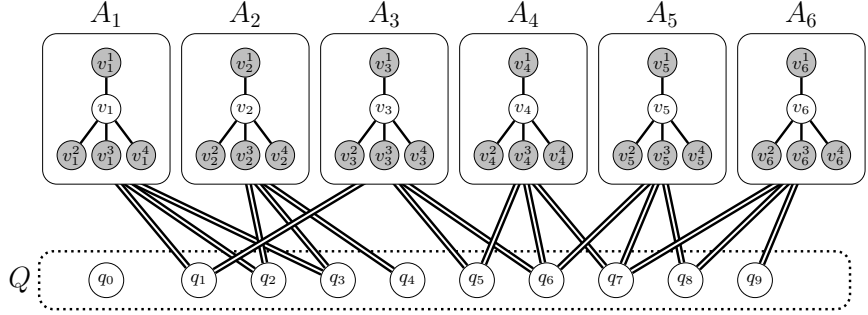
\begin{figure}[t]
    \centering
    \scalebox{0.77}{\input{figures/X3CtoVCDchordal}}
    \caption{Construction of an instance $(G,S,b)$ of {\VCD} from an instance $(U,\mathcal{X})$ of {\ECt}, where $U=\{1,\ldots,9\}$ and $\mathcal{X}=\{X_1=\{1,2,3\}, X_2=\{2,3,4\}, X_3=\{1,5,6\}, X_4=\{5,6,7\}, X_5=\{6,7,8\}, X_6=\{7,8,9\}\}$.  
    The vertices $\cliq_0,\ldots,\cliq_9$ inside the dotted box form a clique of $G$.  
    Each double line between a vertex $\cliq_i$ and a solid box labeled $A_j$ represents all edges between $\cliq_i$ and the vertices in $A_j$.  
    The vertices marked in gray constitute the initial configuration $S$, and $b=9+3=12$.
    }
    \label{fig:X3CoVCDchordal}
\end{figure}

Let $(U,\mathcal{X})$ be an instance of {\ECt}, where the ground set $U = \{u_1, \ldots, u_{3n}\}$ and the family $\mathcal{X}=\{X_1,\ldots,X_m\}$.  
We construct a corresponding instance $(G, S, b)$ of {\VCD} as follows (see \Cref{fig:X3CoVCDchordal} for an illustration).

Let $\Clique=\{\cliq_0,\dots,\cliq_{3n}\}$ be a set of $3n+1$ vertices. 
For each $i \in [3n]$, the vertex $\cliq_i$ corresponds to the element $u_i \in U$.
Next, let $I=\{v_1,\ldots,v_m\}$ be a set of $m$ vertices, where each $v_i$ corresponds to the set $X_i\in \mathcal{X}$.  
We then define a split graph $G'$ with $V(G')=\Clique\cup I$ and $E(G')=\{\cliq\cliq': \cliq,\cliq'\in \Clique\}\cup \{v_i\cliq_j: v_i\in I, \cliq_j\in \Clique, u_j\in S_i\}$.
Since $I$ forms an independent set and $\Clique$ forms a clique in $G'$, it follows that $G'$ is indeed a split graph.

We construct the graph $G$ from $G'$ as follows: for each $v_i\in I$, add the  four vertices $v_i^1,v_i^2,v_i^3,v_i^4$ so that $N(v_i^j)=\{v_i\}\cup(N(v_i)\cap \Clique)$ for each $j\in\{1,2,3,4\}$ in $G$.
Note that each $v_i^j$ is a true twin\footnote{Two vertices $v,v'$ are \emph{true twins} if $N[v]=N[v']$.} of $v_i$ in $G$.
That is, the induced subgraph $G[\{v_i,v_i^1,v_i^2,v_i^3,v_i^4\}]$ forms a star rooted at $v_i$ with leaves $v_i^1,v_i^2,v_i^3,v_i^4$, and all these vertices share the same neighbors in $\Clique$.  
We denote this star by $A_i$.  
Finally, we set $S=\{v_i^1,v_i^2,v_i^3,v_i^4: i\in [m]\}$ and $b=3n+n=4n$.

We observe that the constructed graph $G$ is chordal, since $G$ is obtained from a split graph $G'$ by adding true twins, and split graphs preserve chordality under the operation of adding true twins.
Clearly, the instance $(G, S, b)$ for {\VCD} is constructed in polynomial time.
To complete the polynomial-time reduction, we prove the following lemma. 
The proof can be found in \Cref{appx_subsec:NPcomp_chordal_correctness}.
\begin{lemma}[$\ast$]\label{lem:NPcomp_chordal_correctness}
    An instance $(U,\mathcal{X})$ of {\ECt} is a yes-instance if and only if an instance $(G, S, b)$ of {\VCD} is a yes-instance.
\end{lemma}
This completes the proof of \Cref{thm:VCD_NPcomp_chordal}.

By the equivalence between {\VCD} and {\ISD}, we immediately obtain the $\NP$-hardness of {\ISD} on chordal graphs.
For {\FVSD}, given an instance of {\VCD} on a chordal graph $G$, we construct an equivalent instance of {\FVSD} by replacing each edge $uv$ of $G$ with a triangle on vertices $u, v,$ and $e_{uv}$.  
The resulting graph remains chordal.
Thus, we have the following corollary.

\begin{corollary}\label{cor:FVSD_NPcomp_chordal}
    {\ISD} and {\FVSD} are $\NP$-complete for chordal graphs.
\end{corollary}

Next, we observe the $\NP$-completeness of {\VCD}, {\ISD}, and {\FVSD} on graphs of diameter~$2$.
Recall that if the budget $b$ equals $k \cdot \mathsf{diam}(G)$, then the problem {\PiD} is equivalent to the underlying problem ${\rm \Pi}$.
Therefore, to prove $\NP$-completeness for {\VCD}, {\ISD}, and {\FVSD}, it suffices to show the $\NP$-completeness of the corresponding underlying problems {\VC}, {\IS}, and {\FVS} on graphs of diameter~$2$.
The details can be found in \Cref{appx_subsec:VC_IS_FVS_NPcomp_diam}.
\begin{lemma}[$\ast$]\label{lem:VC_IS_FVS_NPcomp_diam}
    {\VC}, {\IS}, and {\FVS} are $\NP$-complete on graphs of diameter $2$.
\end{lemma}

The following \Cref{cor:VCD_ISD_FVSD_NPcomp_diam} is immediately obtained from \Cref{lem:VC_IS_FVS_NPcomp_diam}.
\begin{corollary}\label{cor:VCD_ISD_FVSD_NPcomp_diam}
    {\VCD}, {\ISD}, and {\FVSD} are $\NP$-complete on graphs of diameter $2$.
\end{corollary}

\subsection{Polynomial-time Algorithms}\label{subsec:polytime_algo}
In this subsection, we show that {\VCD}, {\ISD}, and {\FVSD} can be solved in polynomial time on split graphs.  
We first focus on {\VCD} and then observe that similar results hold for {\ISD} and {\FVSD}.  
Note that split graphs are a subclass of both chordal graphs and graphs of diameter at most $3$.
The details can be found in \Cref{appx_subsec:VCD_split}.

\begin{theorem}[$\ast$]\label{thm:VCD_split}
    {\VCD} can be solved in $O(n^4)$ time on split graphs, where $n$ is the number of vertices in an input graph.
\end{theorem}

To show \Cref{thm:VCD_split}, we enumerate all minimal vertex covers in a given split graph in polynomial time, and then reduce our problem to the problem of finding a minimum-weight matching, in the same manner as \Cref{subsec:FVS_FPT_k}.

We remark that this approach is universal: for any vertex subset problem ${\rm \Pi}$, if one can enumerate candidate solutions of ${\rm \Pi}$ or construct a data structure that encodes such information (e.g., compact representations), {\PiD} can be solved and its running time mainly depends on the enumeration of solutions.
In particular, we can observe that maximal independent sets and minimal feedback vertex sets in a split graph can be enumerated in polynomial time.
Therefore, we have the following corollary.

\begin{corollary}\label{cor:ISD_FVSD_polytime_split}
    {\ISD} and {\FVSD} can be solved in polynomial time on split graphs.
\end{corollary}

We also emphasize that \Cref{thm:VCD_split} and \Cref{cor:ISD_FVSD_polytime_split} can be extended to any graph class in which minimal (maximal for {\ISD}) solutions can be enumerated in polynomial time.

%% file: figures/X3CtoVCDchordal.tex
\begin{tikzpicture}[scale=0.8]

\node[draw, very thick, dotted, rectangle, rounded corners=7pt, minimum width=14cm, minimum height=1cm, label=left:{\Large $\Clique$}] (S1) at (5.3+0.5,0.3) {};

\node[draw=black, circle, minimum size=6mm, inner sep=0pt] (c0) at (-1.5,0.3){$\cliq_0$};
\node[draw=black, circle, minimum size=6mm, inner sep=0pt] (c1) at (0+0.5,0.3){$\cliq_1$};
\node[draw=black, circle, minimum size=6mm, inner sep=0pt] (c2) at (1.5+0.5,0.3){$\cliq_2$};
\node[draw=black, circle, minimum size=6mm, inner sep=0pt] (c3) at (3+0.5,0.3){$\cliq_3$};
\node[draw=black, circle, minimum size=6mm, inner sep=0pt] (c4) at (4.5+0.5,0.3){$\cliq_4$};
\node[draw=black, circle, minimum size=6mm, inner sep=0pt] (c5) at (6+0.5,0.3){$\cliq_5$};
\node[draw=black, circle, minimum size=6mm, inner sep=0pt] (c6) at (7.5+0.5,0.3){$\cliq_6$};
\node[draw=black, circle, minimum size=6mm, inner sep=0pt] (c7) at (9+0.5,0.3){$\cliq_7$};
\node[draw=black, circle, minimum size=6mm, inner sep=0pt] (c8) at (10.5+0.5,0.3){$\cliq_8$};
\node[draw=black, circle, minimum size=6mm, inner sep=0pt] (c9) at (12+0.5,0.3){$\cliq_9$};

\draw[very thick, double, double distance = 1pt] (-1.5,2.7)--(c1);
\draw[very thick, double, double distance = 1pt] (-1.5,2.7)--(c2);
\draw[very thick, double, double distance = 1pt] (-1.5,2.7)--(c3);

\draw[very thick, double, double distance = 1pt] (1.5,2.7)--(c2);
\draw[very thick, double, double distance = 1pt] (1.5,2.7)--(c3);
\draw[very thick, double, double distance = 1pt] (1.5,2.7)--(c4);

\draw[very thick, double, double distance = 1pt] (4.5,2.7)--(c1);
\draw[very thick, double, double distance = 1pt] (4.5,2.7)--(c5);
\draw[very thick, double, double distance = 1pt] (4.5,2.7)--(c6);

\draw[very thick, double, double distance = 1pt] (7.5,2.7)--(c5);
\draw[very thick, double, double distance = 1pt] (7.5,2.7)--(c6);
\draw[very thick, double, double distance = 1pt] (7.5,2.7)--(c7);

\draw[very thick, double, double distance = 1pt] (10.5,2.7)--(c6);
\draw[very thick, double, double distance = 1pt] (10.5,2.7)--(c7);
\draw[very thick, double, double distance = 1pt] (10.5,2.7)--(c8);

\draw[very thick, double, double distance = 1pt] (13.5,2.7)--(c7);
\draw[very thick, double, double distance = 1pt] (13.5,2.7)--(c8);
\draw[very thick, double, double distance = 1pt] (13.5,2.7)--(c9);

\node[draw, fill=white!100, rectangle, rounded corners=7pt, minimum width=2.2cm, minimum height=2.6cm, label=above:{\Large $A_1$}] (S1) at (-1.5,4) {};
\node[draw=black, circle, minimum size=5mm, inner sep=0pt] (v1) at (-1.5,4){$v_1$};
\node[draw=black, fill=gray!50, circle, minimum size=5mm, inner sep=0pt] (v11) at (-1.5,5){$v_1^1$};
\node[draw=black, fill=gray!50, circle, circle, minimum size=5mm, inner sep=0pt] (v12) at (-2.25,3){$v_1^2$};
\node[draw=black, fill=gray!50, circle, circle, minimum size=5mm, inner sep=0pt] (v13) at (-1.5,3){$v_1^3$};
\node[draw=black, fill=gray!50, circle, circle, minimum size=5mm, inner sep=0pt] (v14) at (-0.75,3){$v_1^4$};
\draw[very thick] (v1)--(v11);
\draw[very thick] (v1)--(v12);
\draw[very thick] (v1)--(v13);
\draw[very thick] (v1)--(v14);

\node[draw, fill=white!100, rectangle, rounded corners=7pt, minimum width=2.2cm, minimum height=2.6cm, label=above:{\Large $A_2$}] (S2) at (1.5,4) {};
\node[draw=black, circle, minimum size=5mm, inner sep=0pt] (v2) at (1.5,4){$v_2$};
\node[draw=black, fill=gray!50, circle, circle, minimum size=5mm, inner sep=0pt] (v21) at (1.5,5){$v_2^1$};
\node[draw=black, fill=gray!50, circle, circle, minimum size=5mm, inner sep=0pt] (v22) at (0.75,3){$v_2^2$};
\node[draw=black, fill=gray!50, circle, circle, minimum size=5mm, inner sep=0pt] (v23) at (1.5,3){$v_2^3$};
\node[draw=black, fill=gray!50, circle, circle, minimum size=5mm, inner sep=0pt] (v24) at (2.25,3){$v_2^4$};
\draw[very thick] (v2)--(v21);
\draw[very thick] (v2)--(v22);
\draw[very thick] (v2)--(v23);
\draw[very thick] (v2)--(v24);

\node[draw, fill=white!100, rectangle, rounded corners=7pt, minimum width=2.2cm, minimum height=2.6cm, label=above:{\Large $A_3$}] (S3) at (4.5,4) {};
\node[draw=black, circle, minimum size=5mm, inner sep=0pt] (v3) at (4.5,4){$v_3$};
\node[draw=black, fill=gray!50, circle, circle, minimum size=5mm, inner sep=0pt] (v31) at (4.5,5){$v_3^1$};
\node[draw=black, fill=gray!50, circle, circle, minimum size=5mm, inner sep=0pt] (v32) at (3.75,3){$v_3^2$};
\node[draw=black, fill=gray!50, circle, circle, minimum size=5mm, inner sep=0pt] (v33) at (4.5,3){$v_3^3$};
\node[draw=black, fill=gray!50, circle, circle, minimum size=5mm, inner sep=0pt] (v34) at (5.25,3){$v_3^4$};
\draw[very thick] (v3)--(v31);
\draw[very thick] (v3)--(v32);
\draw[very thick] (v3)--(v33);
\draw[very thick] (v3)--(v34);

\node[draw, fill=white!100, rectangle, rounded corners=7pt, minimum width=2.2cm, minimum height=2.6cm, label=above:{\Large $A_4$}] (S4) at (7.5,4) {};
\node[draw=black, circle, minimum size=5mm, inner sep=0pt] (v4) at (7.5,4){$v_4$};
\node[draw=black, fill=gray!50, circle, circle, minimum size=5mm, inner sep=0pt] (v41) at (7.5,5){$v_4^1$};
\node[draw=black, fill=gray!50, circle, circle, minimum size=5mm, inner sep=0pt] (v42) at (6.75,3){$v_4^2$};
\node[draw=black, fill=gray!50, circle, circle, minimum size=5mm, inner sep=0pt] (v43) at (7.5,3){$v_4^3$};
\node[draw=black, fill=gray!50, circle, circle, minimum size=5mm, inner sep=0pt] (v44) at (8.25,3){$v_4^4$};
\draw[very thick] (v4)--(v41);
\draw[very thick] (v4)--(v42);
\draw[very thick] (v4)--(v43);
\draw[very thick] (v4)--(v44);

\node[draw, fill=white!100, rectangle, rounded corners=7pt, minimum width=2.2cm, minimum height=2.6cm, label=above:{\Large $A_5$}] (S5) at (10.5,4) {};
\node[draw=black, circle, minimum size=5mm, inner sep=0pt] (v5) at (10.5,4){$v_5$};
\node[draw=black, fill=gray!50, circle, circle, minimum size=5mm, inner sep=0pt] (v51) at (10.5,5){$v_5^1$};
\node[draw=black, fill=gray!50, circle, circle, minimum size=5mm, inner sep=0pt] (v52) at (9.75,3){$v_5^2$};
\node[draw=black, fill=gray!50, circle, circle, minimum size=5mm, inner sep=0pt] (v53) at (10.5,3){$v_5^3$};
\node[draw=black, fill=gray!50, circle, circle, minimum size=5mm, inner sep=0pt] (v54) at (11.25,3){$v_5^4$};
\draw[very thick] (v5)--(v51);
\draw[very thick] (v5)--(v52);
\draw[very thick] (v5)--(v53);
\draw[very thick] (v5)--(v54);

\node[draw, fill=white!100, rectangle, rounded corners=7pt, minimum width=2.2cm, minimum height=2.6cm, label=above:{\Large $A_6$}] (S6) at (13.5,4) {};
\node[draw=black, circle, minimum size=5mm, inner sep=0pt] (v6) at (13.5,4){$v_6$};
\node[draw=black, fill=gray!50, circle, circle, minimum size=5mm, inner sep=0pt] (v61) at (13.5,5){$v_6^1$};
\node[draw=black, fill=gray!50, circle, circle, minimum size=5mm, inner sep=0pt] (v62) at (12.75,3){$v_6^2$};
\node[draw=black, fill=gray!50, circle, circle, minimum size=5mm, inner sep=0pt] (v63) at (13.5,3){$v_6^3$};
\node[draw=black, fill=gray!50, circle, circle, minimum size=5mm, inner sep=0pt] (v64) at (14.25,3){$v_6^4$};
\draw[very thick] (v6)--(v61);
\draw[very thick] (v6)--(v62);
\draw[very thick] (v6)--(v63);
\draw[very thick] (v6)--(v64);

\end{tikzpicture}

%% file: 05_conclusion.tex
\section{Conclusion}
In this paper, we investigated the complexity of {\VCD}, {\ISD}, {\DSD}, and {\FVSD} from the perspective of parameterization and graph classes.
We gave a general scheme for solving {\PiD} for various vertex-subset problems ${\rm \Pi}$ in \Cref{sec: FPT/cw} and \Cref{subsec:polytime_algo}.
A promising direction for future research is to identify graph classes for which {\PiD} is $\NP$-complete, while the underlying problem $\mathrm{\Pi}$ is polynomial-time solvable, as demonstrated for chordal graphs.
For interval graphs, which are a subclass of chordal graphs, it is still open whether {\VCD}, {\ISD}, {\DSD}, and {\FVSD} are solvable in polynomial time.
Furthermore, it remains open whether there exists an FPT algorithm for {\FVSD} parameterized by feedback vertex set number.

\subsubsection*{Acknowledgements.}
We thank Akira Suzuki and Xiao Zhou for fruitful discussions.
We are also grateful to the anonymous reviewers for their constructive comments, which improved the presentation.

%% file: 03_Parameter_appendix.tex
\section{Omitted Proofs in \Cref{sec:parameterized}}
\subsection{Proof of \Cref{lem:solsizeFPT}}\label{appx_subsec:solsizeFPT}

We first prove that ``$(\mathrm{1})\Rightarrow (\mathrm{2})$''. 
Suppose that there exists a matching $M$ in $H_{\mathcal{\repres}}$ with total weight at most $b$. 

Let $F'=\{v_1,v_2,\ldots, v_{|\mathcal{\repres}|}\}$ be a minimal feedback vertex set where, for each $i \in [|\mathcal{\repres}|]$, $v_i = \argmin_{v \in \repres_i} \dist(u_i,v)$, that is, $v_i$ is the vertex in $\repres_i$ closest to $u_i$.
We then set $F=F'\cup \{u\in S\colon u \text{ is not matched in }M\}$.
Note that $F$ is a feedback vertex set with size, as containing $F'$.

We now claim that $F$ is reachable in $b$ steps from $S$.
To show this, we construct a sequence of token moves with length at most $b$, as follows:
For each $i \in [|\mathcal{\repres}|]$, move the token $t_i$ initially placed on $u_i$ along a shortest $u_iv_i$-path. 
Since the weight of the edge $(u_i,\repres_i)$ in $H_\mathcal{\repres}$ equals $\dist(u_i,v_i)$, we obtain $\sum_{i\in [|\mathcal{\repres}|]} \dist (u_i,v_i) \le b$.

These token moves may involve another token already occupying a vertex on the path.
Consider the case where a token $t'$ is initially placed on a vertex $u'$ that lies on the shortest $u_iv_i$-path for some $i$, with $t' \neq t_i$. 
If multiple such tokens exist, let $t'$ denote the one closest to $u_i$ along this shortest path. 
Here, we claim that $t'$ must correspond to some token $t_j$ with index $j\in [|\mathcal{\repres}|]$.
To prove this, suppose that $t'$ does not correspond to any token $t_j$ with index $j \in [|\mathcal{\repres}|]$, i.e., $t'$ is a token that does not move in the constructed sequence. 
Since $\dist(u_i,v_i)\ge \dist(u',v_i)$, we can replace the matching edge $(u_i,v_i)$ with $(u',v_i)$, thereby obtaining $M'= (M\setminus \{(u_i,v_i)\})\cup \{(u',v_i)\}$, whose total weight is strictly smaller than that of $M$.
This contradicts the minimality of $M$.

Thus, we can assume that $t'$ can be expressed as $t_j$ for some $j\in [|\mathcal{Y}|]\setminus \{i\}$.
If $t_i$ and $t_j$ are not placed on adjacent vertices, then we simply move $t_i$ toward $v_i$.
Otherwise, i.e., $t_i$ and $t_j$ are placed on adjacent vertices, 
we swap the indices of $t_i$ and $t_j$, and continue moving the newly designated $t_i$ toward $v_j$. 
A new token $t_j$ is moved to $s'$ at the step when $s'$ is not occupied by any token.
Note that $t_j$ cannot reappear on the shortest $u_iv_i$-path before $t_i$ reaches $v_i$; otherwise, the $u_iv_i$-path would contain a cycle, a contradiction. 
By repeating the above swapping procedure, $t_i$ eventually reaches $v_i$. 
This process does not increase the number of token moves, hence the total length of the sequence remains at most $b$.
Therefore, if there is a matching in $H_{\mathcal{\repres}}$ of weight at most $b$, then there exists a feedback vertex set $F$ that is reachable from $S$ within at most $b$ steps.

We next prove that ``$(\mathrm{2})\Rightarrow (\mathrm{1})$''.
Suppose that there exists a feedback vertex set $F$ such that a subset $F'\subseteq F$ is represented by $\mathcal{\repres}$, and consider the sequence of configurations of length at most $b$.
We show that the corresponding bipartite graph $H_\mathcal{\repres}$ admits a matching of weight at most $b$ that every vertex in $\mathcal{\repres}$ is saturated.

Let $\mathcal{P}=\{P_1,P_2,\ldots, P_k\}$ be the set of walks, where each walk $P_i$ describes the moves of a token $t_i$ from its initial position $u_i$ to its target position $g_i \in F$.  
By construction, the length of each walk does not exceed the length of the shortest $u_ig_i$-path, thus the total length $\sum_{i\in [k]} \dist(u_i,g_i)$ is at most $b$. 
Now consider the set $\mathcal{P}'\subseteq \mathcal{P}$ of walks that correspond to the tokens placed on $F'$ in the target configuration.

Since $F'$ is a minimal feedback vertex set of $G$ and $F'$ is represented by $\mathcal{\repres}$, there is a matching $M'=\{(u_1,\repres_1),(u_1,\repres_2),\ldots ,(u_{|\mathcal{\repres}|},\repres_{|\mathcal{\repres}|})\}$ of $H_\mathcal{\repres}$ such that each $\repres_i$ contains the target vertex $g_i$.
Moreover, since each edge $(u_i,\repres_i)$ in $H_\mathcal{\repres}$ is assigned the weight $\min_{c \in \repres_i}\dist(u_i,c)$, it follows that the weight of $M'$ is at most $b$.  
This completes the proof.

%% file: 04_GraphClass_appendix.tex
\section{Omitted Proofs in \Cref{sec:NPcomp}}
\subsection{Proof of \Cref{lem:NPcomp_chordal_correctness}}\label{appx_subsec:NPcomp_chordal_correctness}
    We first show the ``only if'' direction.
    Suppose that $(U,\mathcal{X})$ is a yes-instance of {\ECt}, and hence there exists a subfamily $\mathcal{X}' \subseteq \mathcal{X}$ of exactly $n$ sets whose union forms $U$.
    
    From this subfamily, we construct a vertex cover $\Cover$ that can be reached from $S$ in exactly $b$ steps.
    For each $X_i \in \mathcal{X}'$ containing elements $u_{j_1}, u_{j_2}, u_{j_3}$ with $j_1,j_2,j_3 \in [3n]$, we move the three tokens initially placed on $v_i^2,v_i^3,v_i^4$ to the corresponding vertices $\cliq_{j_1},\cliq_{j_2},\cliq_{j_3}$ of $\Clique$, and slide the token on $v_i^1$ to $v_i$.
    Since this procedure requires four steps for each $X_i$, the total number of steps is $4n = b$.

    By construction, $\Cover$ contains all vertices $\cliq_1,\ldots,\cliq_{3n}$ together with either the root or all leaves of each star $A_i$, and hence $\Cover$ is indeed a vertex cover of $G$.
    Recall that $\cliq_0\in \Clique$ is not adjacent to any vertex of the star $A_i$ for every $i\in[m]$.
    Thus, $(G,S,b)$ is a yes-instance of {\VCD}.

    We next show the ``if'' direction.
    Suppose that $(G,S,b)$ is a yes-instance of {\VCD}, and hence there exists a vertex cover $\Cover$ of $G$ that is reachable from $S$ in $b$ steps.

    Since $\Clique$ is a clique of $G$ with $3n+1$ vertices, any vertex cover of $G$ must contain at least $3n$ vertices from $\Clique$.
    As the initial configuration $S$ places no tokens on $\Clique$, moving tokens there requires at least $3n$ steps.
    
    Moreover, to cover the edges of each star $A_i$ for $i \in [m]$, at least one token must be placed on a vertex of $A_i$.
    Consequently, at least $3n/3=n$ stars must contribute at most three tokens to $\Clique$, which forces the roots of these $n$ stars to be included in $\Cover$.
    Moving tokens to these roots requires at least $n$ additional steps.

    Consequently, exactly $n$ stars $A_{i_1}, \dots, A_{i_n}$ each contribute three tokens to the vertices in $\Clique$, while the remaining token in each star is moved to its root.
    Therefore, the sets $X_{i_1}, \dots, X_{i_n}$ together cover all elements of $U$, and hence $(U,\mathcal{X})$ is a yes-instance of {\ECt}.

\subsection{Proof of \Cref{lem:VC_IS_FVS_NPcomp_diam}}\label{appx_subsec:VC_IS_FVS_NPcomp_diam}

    We first observe that {\VC} remains $\NP$-complete on graphs of diameter $2$.  
    Given an instance $(G,k)$ of {\VC}, construct $(G',k+1)$ by adding a universal vertex $u$ adjacent to every vertex in $G$.  
    Then, we can see that $(G,k)$ is a yes-instance if and only if $(G',k+1)$ is.  
    Thus, \prb{Vertex Cover} is $\NP$-complete for graphs of diameter at most $2$.
    By the equivalence between {\VC} and {\IS}, {\IS} is also $\NP$-complete for graphs of diameter at most $2$.

    Next, we show that the claim for {\FVS} is reduced from {\FVS} on general graphs.
    Given an instance of {\FVS} $(G,k)$, construct $(G',k+1)$ with 
    \begin{align*}
    &V(G') = V(G) \cup \{s\} \cup \{v_i^1,v_i^2 : i \in [k+2]\}, \\ 
    &E(G') = E(G) \cup \{sv : v \in V(G)\} \cup \{sv_i^1, sv_i^2, v_i^1v_i^2 : i \in [k+2]\}.  
    \end{align*}

    Each triple $s, v_i^1, v_i^2$ forms a triangle, and $s$ is a universal vertex, so $G'$ has diameter at most $2$. 

    It is straightforward to verify that $(G,k)$ is a yes-instance if and only if $(G',k+1)$ is.  
    Specifically, any feedback vertex set of $G'$ of size at most $k+1$ must include $s$, and the remaining $k$ vertices form a feedback vertex set of $G$.

\subsection{Proof of \Cref{thm:VCD_split}}\label{appx_subsec:VCD_split}
    We employ a technique from the algorithm constructed by Fellows et al.\ for {\VCD} parameterized by $k$~\cite{SolDicov:FellowsGMMRRSS23}, as well as from our FPT algorithm for {\FVSD} parameterized by $k$ presented in \Cref{subsec:FVS_FPT_k}.

    Let $G=(V,E)$ be a split graph that can be partitioned into a clique $\Clique$ and an independent set $I$.   
    We assume that each $v \in \Clique$ has a neighbor in $I$; otherwise, if $u \in \Clique$ has no neighbor in $I$, then we can form another partition with a clique $C \setminus \{u\}$ and an independent set $I \cup \{u\}$.
    First, we observe that only a small number of token placements $\Cover$ need to be considered.  
    In particular, any minimal vertex cover must place at least $|\Clique|-1$ tokens on $\Clique$; otherwise, $\Clique \setminus \Cover$ contains an uncovered edge.  
    If all $|\Clique|$ tokens are placed on $\Clique$, then $\Cover = \Clique$.  
    If $|\Clique|-1$ tokens are placed on $\Clique$, there are exactly $|\Clique|$ choices: for the vertex $v \in \Clique$ not in $\Cover$, its neighborhood must be included in $\Cover$, giving $\Cover = (\Clique \setminus \{v\}) \cup N(v)$.  
    Thus, the number of minimal vertex covers is exactly $|\Clique| + 1$, and enumeration can be done in $O(n^2)$ time.

    Next, for each minimal vertex cover $\Cover$, we construct a complete weighted bipartite graph $H_\Cover$ with bipartition $(S,\Cover)$, where the weight of the edge $uv$ is $\dist(u,v)$ in $G$.  
    We then compute a minimum-weight maximum matching in $H_\Cover$ that matches all vertices in $\Cover$.  
    If such a matching exists with total weight at most $b$, then $\Cover$ is reachable from $S$ within $b$ steps, and the algorithm outputs $\yes$; otherwise, we proceed to the next cover. 
    If no cover satisfies the condition, the algorithm returns $\no$.  
    Correctness follows the same argument as in \Cref{lem:solsizeFPT}.

    For the running time, precomputing all-pairs distances takes $O(n^3)$ time.  
    Constructing $H_\Cover$ and solving the problem of finding a minimum-weight maximum matching using the Hungarian algorithm takes $O(n^2 + n^3) = O(n^3)$ for each minimal vertex cover.  
    Since there are $|\Clique| + 1 = O(n)$ minimal covers, the overall running time is $O(n^3 + n \cdot n^3) = O(n^4)$.